\documentstyle[preprint,eqsecnum,aps]{revtex}
\begin{document}


\begin{titlepage}

\title{A Hybrid Decomposition Parallel Implementation\\
of the Car-Parrinello Method}

\author{James Wiggs and Hannes J\'onsson${^*}$}
\address{Department of Chemistry, BG-10\\
University of Washington\\
Seattle, WA 98195}

\date{\today}

\maketitle

\vskip 0.5 true cm
\centerline{{\it Computer Physics Communications} (in Press)}
\vskip 0.5 true cm

\begin{abstract}

   We have developed a flexible hybrid decomposition parallel implementation of
the first-principles molecular dynamics algorithm of Car and Parrinello.  The
code allows the problem to be decomposed either spatially, over the electronic
orbitals, or any combination of the two.  Performance statistics for 32, 64,
128 and 512 Si atom runs on the Touchstone Delta and Intel Paragon parallel
supercomputers and comparison with the performance of an optimized code running
the smaller systems on the Cray Y-MP and C90 are presented.

\end{abstract}


\end{titlepage}


\section{Introduction}
\label{sec:level1}

   The ab-initio molecular dynamics technique of Car and Parrinello
\cite{CP85,CP88_1} has become a valuable method for studying condensed matter
structure and dynamics, in particular liquids
\cite{CP88_2,Stich89,Stich91,Hohl90}, surfaces
\cite{Brocks91,Payne93,Brommer92}, and clusters
\cite{Andreoni87,Hohl87,Ballone88}.  In a Car-Parrinello (CP) simulation the
electron density of the ground electronic state is calculated within the Local
Density Approximation (LDA) of Density Functional Theory (DFT), and is used to
calculate the forces acting on the ions.  The electronic orbitals are expanded
in a plane-wave basis set; a classical Lagrangian linking the ionic coordinates
with the expansion coefficients is then used to generate a coupled set of
equations of motion describing the extended electron-ion system.  The motion of
the electrons can be often adjusted so that they follow the motion of the ions
adiabatically, remaining close to the Born-Oppenheimer ground state as the
system evolves.  Thus the ions move according to ab-initio forces determined
directly from the electronic ground state at each time step, rather than from
an empirical potential.  As such, the CP algorithm overcomes many of the
limitations of standard empirical-potential approaches, such as transferability
of potentials and, furthermore, provides direct information about the
electronic structure.  However, CP simulations are computationally demanding
and systems larger than ${\sim 100}$ atoms can not be simulated in a reasonable
amount of time on traditional vector supercomputers.  Furthermore, the memory
requirements for simulations of large systems easily exceed the available
memory on shared-memory supercomputers.

   Many systems of interest require simulations of $10^3$ atoms as is commonly
done in molecular dynamics simulations with empirical potentials.  In order to
study larger systems with the CP approach it is necessary to take advantage of
the computing power and memory available on parallel supercomputers.  A
parallel supercomputer can offer a considerable increase in performance over
traditional vector supercomputers by replacing the small number of processors
sharing a single memory space (e.g. the Cray C-90) with a large number of
computing nodes, each consisting of a slower -- but much less expensive --
processor with its own memory.  The nodes have some means of communicating data
with one another to work cooperatively when solving the problem.  An efficient
use of parallel computers requires that a significant fraction of the
computation can be done in an arbitrary order, such that tasks can be done
simultaneously on the individual nodes without excessive data communication
between the nodes.

   Several groups have implemented the CP algorithm
\cite{Stich92,Brommer93,Clarke92,Wiggs94} or similar plane-wave
electronic-structure calculations \cite{Nelson93} on parallel supercomputers.
A large fraction of the computation in such calculations involves fast Fourier
transforms (FFT).  Most groups \cite{Stich92,Brommer93,Clarke92,Nelson93} have
used a spatial decomposition of the problem where each node was made
responsible for calculations on a subset of the plane-wave coefficients used to
describe each orbital, basically implementing a parallel FFT.  This allows
efficient implementation of calculations involving several different orbitals,
most importantly the orthonormalization of the orbitals and the non-local
portion of the Hamiltonian.  However, it is quite difficult to implement
domain-decomposition multi-dimensional FFTs efficiently, particularly on
parallel computers with low degrees of connectivity such as the mesh-connection
\cite{Gupta93,Marinescu92,Angelopoulos93} due to the communication
requirements.  This seriously reduces the efficiency when calculating the
electronic density and the action of the local part of the Hamiltonian.
Alternatively, we chose an orbital decomposition where each node was made
responsible for {\it all} the expansion coefficients for a subset of the {\it
orbitals} \cite{Wiggs94}.  This approach has advantages and disadvantages when
compared with the spatial decomposition.  Much of the computation involves
independent operations on the orbitals such as the FFT, and the orbital
decomposition makes it possible to do them fully in parallel without any
communication between nodes.  One disadvantage of the orbital decomposition is
that the orthonormalization requires extensive communication between nodes;
another disadvantage of the orbital decomposition when applied to large
problems is the requirement that each node set aside memory for the entire
3-dimensional FFT lattice, rather than storing only a small portion of it.

   As more and more nodes are applied to the calculation, a point of
diminishing returns is reached in both cases.  There are irreducible minimum
amounts of communication, redundant computation done on each node to avoid
communication, synchronization, and load-balancing problems which all act
together to limit the efficiency of any parallel implementation.  In order to
apply the CP algorithm to large problems, and to make optimal use of parallel
computers, we have developed a code which can use a {\it combination} of the
orbital and spatial decomposition.  This relieves the memory limitations of the
orbital decomposition, and makes it possible to balance out the losses in
efficiency for different parts of the computation so as to get {\it optimal}
speedup for a given number of nodes.  We present here results of tests on this
code on different sized problems.  We find that the optimal decomposition in
every case we studied was neither purely spatial nor purely orbital, but rather
a combination of the two.


\section{The CP Algorithm}
\label{sec:level2}

   The CP algorithm describes a system of interacting ions and electronic
orbitals with the classical Lagrangian \cite{CP85,CP88_1,CP88_2}

\begin{eqnarray}
L = & \sum^N_n  \ \mu f_n \int_\Omega  d{\vec r} \ | \dot \psi_n({\vec r})|^2 \
+ \ {1 \over 2} \sum_I \ M_I {\dot {\vec R}_I}^2 \ - \ E [ \{ {\vec R}_I \} ,\{
\psi_n \} ] \nonumber \\
& \ + \sum_{n,m} \Lambda_{nm} \ \left( \int {d{\vec r} \ \psi^*_n({\vec r})
\psi_m({\vec r}) } - \delta_{nm} \right) \label{eq1}
\end{eqnarray}

\noindent
where the $\{ {\vec R}_I \}$ are the ionic coordinates and the $M_I$ their
masses, dots indicate time derivatives, $\Omega$ is the volume of the
simulation cell, $\int_{\Omega}$ indicates integration over the simulation
cell, $\mu$ is a fictitious mass associated with the time-dependent electronic
orbitals to control the time scale of the electronic motion (typically $\mu <<
M_I$) and $E$ is energy functional of the system within the DFT.  The
$\Lambda_{nm}$ are Lagrange multipliers which impose orthonormality constraints
on the electronic orbitals $\psi_n$:

\begin{equation}
< \psi_n | \psi_m > \ = \int { d{\vec r} \ \psi^*_n({\vec r}) \psi_m({\vec r})
} \ =\  \delta_{nm} \label{eq2}
\end{equation}

\noindent
where the $\delta_{nm}$ is the standard Kronecker delta function, equal to one
when $n=m$ and zero otherwise.  The classical equations of motion are derived
from the Lagrangian in the usual way:

\begin{equation}
\mu \ddot \psi_n({\vec r},t) = - {1 \over 2} { {\delta E} \over {\delta
\psi^*_n({\vec r},t)} } + \sum_m {\Lambda_{nm} \psi_m({\vec r},t) } \label{eq3}
\end{equation}

\noindent
and

\begin{equation}
M_I {\ddot {\vec R}_I} = - {{\partial E} \over {\partial {{\vec R}_I}(t)}}
\label{eq4}
\end{equation}

\noindent
The DFT energy functional \cite{Hohenberg64,KohnSham65} has the form

\begin{eqnarray}
E[ \{ {\vec R}_I \},\{ \psi_n \} ] = & \sum^N_n \ f_n \ {\int d{\vec r} \
\psi^*_n(\vec r) \left[ - {1 \over 2} \nabla^2 \right]\ \psi_n(\vec r) } \ + \
\int { d{\vec r} \ V^{ext}({\vec r}) \ \rho({\vec r}) } \ \nonumber \\
& + \  {1 \over 2} \int { d{\vec r} d{\vec r}^\prime \ { {\rho({\vec r})
\rho({\vec r}^\prime)} \over { | {\vec r} - {\vec r}\prime | } } } \ + \ E^{xc}
[ \rho ] \ + \  {1 \over 2} \sum_{I \neq J} \ { { Z_{I} Z_{J} } \over { | {\vec
R}_I - {\vec R}_{J} | } } \label{eq5}
\end{eqnarray}

\noindent
where $\rho({\vec r})$ is the electron density at $\vec r$, $E^{xc} [ \rho ]$
is the LDA of the exchange-correlation energy for electronic density $\rho$,
$f_n$ is the occupation number of orbital $\psi_n$, $Z_{I}$ is the valence
charge of atom $I$, and $V^{ext}({\vec r})$ is a sum of ionic pseudopotentials.
 We use the angular-momentum dependent, norm-conserving pseudopotentials of
Bachelet, Hamann, and Schl{\" u}ter (BHS) \cite{Bachelet82} in the factorized
form of Kleinman and Bylander \cite{KlByl82} for the nonlocal parts.  The
electron density $\rho({\vec r})$ is expressed in terms of the N orbitals
$\psi_n$

\begin{equation}
\rho(\vec r) \ = \ \sum^N_n \ f_n \ |\psi_n(\vec r)|^2  \label{eq6}
\end{equation}

   If the functional $E[ \{ {\vec R}_I \},\{ \psi_n \}]$ is minimized with
respect to the electronic orbitals for fixed ionic positions, the BO potential
surface for the ions, $\Phi [ \{ {\vec R}_I \} ]$, is obtained.  The equations
of motion derived from Eq. (\ref{eq1}) make it possible to optimize
simultaneously the electronic and ionic degrees of freedom using, for example,
Steepest Descent (SD) minimization.  Furthermore, they allow one to perform
finite-temperature molecular dynamics on the BO potential surface once the
electronic degrees of freedom have been minimized.  Under favorable conditions
the value of $\mu$ can be chosen such that the electronic orbitals can remain
close to the BO surface as the ionic coordinates evolve according to Eq.
(\ref{eq4}).  When doing finite-temperature simulations of metallic systems, it
is often necessary to periodically re-quench the electronic orbitals to the BO
surface by holding the ions fixed and performing SD or Conjugate Gradient (CG)
minimization on the electrons according to Eq. (\ref{eq3}).  The temperature of
the ions can be controlled by any convenient thermostat, such as velocity
scaling, stochastic collisions\cite{Andersen80}, or the Nos{\' e}-Hoover
thermostat \cite{Nose84,Hoover85}.


\section{Numerical Implementation}
\label{sec:level3}

   The CP algorithm is most easily applied by expanding the electronic orbitals
in sums of plane waves:

\begin{equation}
\psi_{n,k}({\vec r}) =  e^{i \vec k \cdot \vec r} \sum_{\vec g} { c^n_{\vec g}
e^{i {\vec g} \cdot {\vec r} } }\label{eq7}
\end{equation}

\noindent
where the ${\vec g}$'s are reciprocal lattice vectors of the simulation cell:

\begin{equation}
{\vec g} = n_x { {2 \pi} \over {a_x} } {\hat x} + n_y { {2 \pi} \over {a_y} }
{\hat y} + n_z { {2 \pi} \over {a_z} } {\hat z} \label{eq8}
\end{equation}

\noindent
where $a_x$, $a_y$, and $a_z$ are the dimensions of the simulation cell, $n_x$,
$n_y$, and $n_z$ can have any integer value and ${\hat x}$, ${\hat y}$, and
${\hat z}$ are the unit vectors in the $x$, $y$, and $z$ directions.  The
expression is simple in this case due to the fact that our simulation cell has
all internal angles equal to 90 degrees.  Other simulation cell symmetries can
be used, but result in a more complex expression for the reciprocal lattice
vector's basis set.  The set of ${\vec g}$ is limited to those whose kinetic
energy $E_{kin} = {1 \over 2} |{\vec g}|^2$ is less than some energy cutoff
$E_{cut}$; a larger value of $E_{cut}$ increases the accuracy of the expansion.
 The number of such ${\vec g}$ is hereafter referred to as $M$.  A typical
value of $E_{cut}$ for a simulation of silicon is about 12.0 Rydbergs ($\simeq
163\ eV$).  In principle, several $\vec k$ vectors need to be included to
sample the first Brillouin zone; however, this becomes less and less important
as the size of the simulation cell is increased.  Since we were primarily
interested in simulating large systems with our parallel code the only k-point
included is $(0,0,0)$, the $\Gamma$ point.  This choice of basis set forces the
orbitals to have the same periodicity as the simulation cell when periodic
boundary conditions are applied.  It has the additional benefit of making the
phase of the wavefunction arbitrary; we can therefore choose it to be real,
which is equivalent to stating that  $c_{-{\vec g}} = c^*_{\vec g}$, which
reduces the required storage for the expansion coefficients by a factor of two.

   The plane wave expansion has the added benefit that certain parts of the
Hamiltonian are very easily calculated in terms of the ${\vec g}$'s, i.e. in
reciprocal space.  While other parts of the calculation are more efficiently
carried out in real space, the plane wave basis makes it possible to switch
quickly from reciprocal space to real space and back using FFTs.  This reduces
the work required to calculate the functional derivatives in Eq. (\ref{eq3})
from  $O( N M^2 )$ to $O( N M log\strut M )$, making the most computationally
expensive portion of the calculation the imposition of orthonormality and the
Kleinman-Bylander nonlocal pseudopotentials, which require $O( N^2 M )$
computation.  Any parallel implementation of the algorithm will have to perform
these parts of the calculation efficiently to achieve good speedup.


\subsection{The pseudopotential calculation}
\label{sec:level31}

   Introduction of pseudopotentials not only reduces the number of electrons
included in the calculation by allowing us to treat only the valence electrons,
it also greatly reduces the size of the basis set required to accurately
describe the wavefunctions and the electron density and potential since it is
not necessary to reproduce the fine structure in the regions of space near the
nuclei.  The Kleinman-Bylander factorized form of the pseudopotentials first
describes the interaction of the valence electrons with the ionic cores as a
sum of ionic pseudopotentials

\begin{equation}
V({\vec r}) = \sum_I { v_{ps}({\vec r} - {\vec R_I}) } \label{eq9}
\end{equation}

\noindent
then breaks these pseudopotentials down further into sums of angular-momentum
dependent potentials

\begin{equation}
v_{ps}({\vec r}) = \sum^\infty_{l=0} { v_l(r) {\hat P}_l } \label{eq10}
\end{equation}

\noindent
where ${\hat P}_l$ projects out the l-th angular momentum.  The assumption is
made that for some $l > {lmax}$,  $v_l(r) = v_{lmax}(r)$.  For most elements,
this approximation is good for ${lmax} = 1 \ or \ 2$.  Since ${\hat P}_l$ is a
complete set, Eq. (\ref{eq10}) can be written as:

\begin{equation}
v_{ps}({\vec r}) = v_{loc}({\vec r}) + \sum^{{lmax} - 1}_{l=0} {\delta v_l(r)
{\hat P}_l } \label{eq11}
\end{equation}

\noindent
with

\begin{equation}
\delta v_l(r) = v_l(r) - v_{lmax}(r) \label{eq12}
\end{equation}

   The Kleinman-Bylander formalism then replaces the sum in Eq. (\ref{eq11})
with a fully nonlocal potential:

\begin{equation}
{\tilde v}_{nl}(r) = \sum_{l,m} { {|\delta v_l \Phi^0_{l,m}> <\Phi^0_{l,m}
\delta v_l|} \over {<\Phi^0_{l,m}|\delta v_l|\Phi^0_{l,m}>} } \label{eq13}
\end{equation}

\noindent
where $\Phi^0_{l,m} = \phi_l Y_{lm}$, with $Y_{lm}$ being the spherical
harmonic and $\phi_l$ the {\it l}-th atomic pseudowavefunction from which the
$v_l$ were originally generated.  In the plane-wave implementation, if we make
the following definitions:

\begin{equation}
u_l(\vec g) = \int  dr \ r^2 j_l(gr) \delta v_l(r) \phi_l(r) \label{eq14}
\end{equation}

\begin{equation}
W_l = \int  dr \ r^2 j_l(r) \delta v_l(r) \phi_l(r)^2 \label{eq15}
\end{equation}

\noindent
then it can be shown that the matrix element of the nonlocal part of the
Hamiltonian in reciprocal space is:

\begin{equation}
<{\vec g}| {\tilde v}_{nl} |{\vec g}^\prime> = {{4 \pi} \over \Omega} \sum_l {{
(2l+1) P_l( cos( \theta_{{\vec g},{\vec g}^\prime} ) ) }\over{W_l}} u_l(\vec g)
u_l({\vec g}^\prime) \label{eq16}
\end{equation}

   The forces on the electronic coefficients due to their interaction with the
ions is:

\begin{equation}
{{\partial {\tilde v}^{l}_{nl}(\vec g) } \over {\partial c^n_{\vec g}}} = 2 {
{\sum_I e^{-i \vec g \cdot {\vec R}_I} u_l(\vec g) F_l^{In}} \over { W_l } }
\label{eq17}
\end{equation}

\noindent
where

\begin{equation}
F_l^{In} = \sum^M_{\vec g} { u_l(\vec g) e^{i \vec g \cdot {\vec R}_I}
c^n_{\vec g} } \label{eq18}
\end{equation}

   Calculating the $ F_l^{In} $ is the most computationally demanding part of
the pseudopotential calculation, requiring $O(N^2 M)$ work.  It should be noted
that implementations of the non-local pseudopotentials calculated in real space
have been suggested \cite{Troullier91,KS91} which would reduce the scaling of
this calculation to $O(N^2)$, but with a large prefactor.  Such techniques
could be efficiently implemented within our hybrid parallel code, reducing the
scaling of this portion of the calculation to $O(N)$ when the number of
processors is increased with the number of atoms, but we have not yet done so.


\subsection{The orthonormalization}
\label{sec:level32}

   The orthonormality of the electronic orbitals may be maintained in two ways:
either by a straightforward technique like Gram-Schmidt (GS) orthonormalization
or by an iterative technique as given by Car and Parrinello \cite{CP88_1} based
upon the more general method for imposing holonomic constraints described by
Ryckaert et. al. \cite{Ryckaert77}.  Both techniques have been implemented in
our code, but the iterative technique is preferred when doing ionic dynamics.

   Applying the functional derivatives in Eq. (\ref{eq3}) to the  $\{ \psi_n
\}$, produces a new, non-orthonormal set $\{ {\bar \psi}_n \}$.  This set can
be brought to orthonormality using the real symmetric matrix ${\bf X} =
{{(\delta t)^2} \over {\mu}} {\bf \Lambda}$.  ${\bf X}$ can be recovered by
first defining the matrices ${\bf A}$ and ${\bf B}$ as

\begin{equation}
A_{nm} = < \bar \psi_n | \bar \psi_m > \label{eq19}
\end{equation}

\begin{equation}
B_{nm} = < \psi_n | \bar \psi_m > \label{eq20}
\end{equation}

\noindent
then making a first order approximation to ${\bf X}$ with:

\begin{equation}
{\bf X}^{(0)} = {1 \over 2}( {\bf I} - {\bf A} ) \label{eq21}
\end{equation}

\noindent
and iterating

\begin{equation}
{\bf X}^{(k)} = {1 \over 2} \left[ {\bf I} - {\bf A} + {\bf X}^{(k-1)}({\bf I}
- {\bf B}) + ({\bf I} - {\bf B})^T{\bf X}^{(k-1)} - {\bf X}^{(k-1)^2} \right]
\label{eq22}
\end{equation}

\noindent
until

\begin{equation}
Max { | X^{(k)}_{nm} - X^{(k-1)}_{nm} | } < \epsilon \label{eq23}
\end{equation}

   The value of $\epsilon$ determines how close to orthonormal the orbitals
remain and how many iterations of Eq. (\ref{eq22}) are required to achieve
orthonormality.  A typical value of $10^{-6}$ generally requires less than four
iterations.

   In a plane wave basis set, with $\Gamma$ point symmetry, it can be shown
that the overlap of two wavefunctions is:

\begin{eqnarray}
< \psi_n | \psi_m > \  = & {1 \over \Omega} \int_{\Omega} { d{\vec r} \
\psi^*_n({\vec r}) \psi_m({\vec r}) }\nonumber \\
\ = & Re[c^n_{\vec 0}] \cdot Re[c^m_{\vec 0}]  + 2 \sum^M_{\vec g \neq \vec 0}
\left( { Re[c^n_{\vec g}] \cdot Re[c^m_{\vec g}] + Im[c^n_{\vec g}] \cdot
Im[c^m_{\vec g}] } \right) \label{eq24}
\end{eqnarray}

\noindent
so that calculating each element ${\bf A}$ and ${\bf B}$ reduces to doing
slightly modified dot product, and thus calculating ${\bf A}$ and ${\bf B}$
requires multiplying one $M \times N$ matrix of expansion coefficients by
another matrix with dimension $N \times M$.  An operation requiring $O(N^2M)$
computation.  It is possible to exploit certain matrix properties, store
intermediate values of some matrices, and define the iteration matrix somewhat
differently, so that the number of multiplications in the iteration loop is
reduced to one.  Since these are multiplications of $N \times N$ matrices,
requiring $O(N^3)$ computation, and $M>>N$, the overall time is reduced, but
only marginally.  The calculations required, in a plane wave basis set, reduce
to:

\begin{eqnarray}
2X^0_{nm} = & \delta_{nm} + Re[\bar c^n_{\vec 0}(t + \delta t)] \cdot Re[\bar
c^m_{\vec 0}(t + \delta t)] - \nonumber \\
& \sum^M_{\vec g} \left( { Re[\bar c^n_{\vec g}(t + \delta t)] \cdot Re[\bar
c^m_{\vec g}(t + \delta t)] + Im[\bar c^n_{\vec g}(t + \delta t)] \cdot Im[\bar
c^m_{\vec g}(t + \delta t)] } \right) \label{eq25}
\end{eqnarray}

\begin{eqnarray}
b_{nm} = & \delta_{nm} + Re[c^n_{\vec 0}(t)] \cdot Re[\bar c^m_{\vec 0}(t)] -
\nonumber \\
& \sum^M_{\vec g} \left( { Re[c^n_{\vec g}(t)] \cdot Re[\bar c^m_{\vec g}(t +
\delta t)] + Im[c^n_{\vec g}(t)] \cdot Im[\bar c^m_{\vec g}(t + \delta t)] }
\right) \label{eq26}
\end{eqnarray}

\begin{equation}
X^{(k)} = X^{(0)}  +  X^{(k-1)}  b +  b^T  X^{(k-1)}  - X^{(k-1)^2}
\label{eq27}
\end{equation}

\begin{equation}
c^n_{\vec g}(t + \delta t) = \bar c^n_{\vec g}(t + \delta t) + \sum^N_{m=1}
X^{(k)}_{nm} c^m_{\vec g}(t) \label{eq28}
\end{equation}


\subsection{Summary of the computational modules}
\label{sec:level33}

   A schematic summary of the evaluation of the force on the electronic
coefficients is given in Table \ref{table1}.   Over 90 \% of the numerical
computation is spent in dealing with the electronic degrees of freedom; it
breaks down into six major tasks. Below is a description of the work involved
in each task.

  {\bf Rhoofr} involves calculation of the total electronic density $\rho({\vec
r})$ in the simulation cell (Eq. \ref{eq6}).

   {\bf Vofrho} uses the total electronic density generated by {\bf rhoofr} to
determine the total local electronic potential as a function of $\rho({\vec
r})$ throughout the simulation cell, including the contribution of the electron
exchange-correlation potential and Hartree interactions (Eq. \ref{eq5}), and
the {\it local} portion of the pseudopotential (Eq. \ref{eq11}).  The
pseudopotential and the Hartree interactions are determined in reciprocal
space; the exchange-correlation potential in real space.

   {\bf Nonlocal} determines the portion of the unconstrained functional
derivatives of the electronic coefficients due to the nonlocal portion of the
pseudopotential, using the Kleinman-Bylander factorized form (Eq. \ref{eq17}).
In addition, when doing ionic dynamics, it calculates the force exerted on the
ions by the electrons, interacting through the nonlocal pseudopotential.

   {\bf Local} determines the portion of the unconstrained functional
derivatives of the electronic coefficients due to the total local potential
calculated in {\bf vofrho}.

   {\bf Loop} updates the sets of electronic coefficients according to these
unconstrained functional derivatives:

\begin{equation}
\bar c^n_{\vec g}(t+\delta t) = - c^n_{\vec g}(t-\delta t) + 2 c^n_{\vec g}(t)
-  {{(\delta t)^2} \over \mu} \ {{\partial E } \over {\partial c^n_{\vec g}}}
\label{eq29}
\end{equation}

\noindent
where $ \bar c^n_{\vec g}(t+\delta t) $ are the new set of expansion
coefficients {\it {before application of the constraint forces}}.

   Orthonormalization is carried out either in {\bf ortho} via calculation and
application of constraint forces, or in {\bf gram} via the simple Gram-Schmidt
procedure.


\section{Parallel Implementation}
\label{sec:level4}

   Table \ref{table1} gives a schematic representation of the CP algorithm.  It
suggests two approaches to the parallel implementation.  It is noted that the
work on each electronic orbital $\psi_n$ is largely independent of the work
done on the other electronic orbitals; this implies that dividing the orbitals
up among the nodes, an orbital decomposition, may be successful, and in fact
this proves to be the case \cite{Wiggs94}.  This type of parallelism is often
referred to as coarse-grain or macro-tasking parallelism; the amount of work
assigned to each node is quite large, and the number of nodes which can be
applied to the problem is limited.  Closer examination suggests that the work
done on each expansion coefficient $c^n_{\vec g}$ is also independent of the
work done on the other coefficients.  For example {\bf loop}, the second and
third subtask of {\bf nonlocal}, and the second subtask in {\bf vofrho}.  This
implies that we might divide up the coefficients among the nodes, an example of
fine-grain or micro-tasking parallelism, the approach usually favored by
parallelizing compilers.  The number of nodes which can be applied to the
problem in this manner is theoretically limited only by the number of
coefficients, but effectively the limit is much smaller due to load balancing
and communications requirements in some parts of the code.  This spatial
decomposition approach has been utilized by several groups
\cite{Stich92,Brommer93,Clarke92,Nelson93}.

   Each approach has advantages and drawbacks.  Figure \ref{fig1} shows the
time spent doing various tasks in the algorithm for a 32 atom Si calculation,
using the pure orbital decomposition.  It shows excellent speedup for {\bf
rhoofr} and application of the local part of the potential to the
wavefunctions, due to the fact that the FFTs are done with no communication,
completely in each processors local memory -- inspection of the actual timing
figures shows essentially 100\% efficiency.  Speedup of the non-local part of
the computation is not as good due to some redundant computation carried out on
each node.  Unfortunately, since each node must do the calculations for all
${\vec g}$, most of the work in {\bf vofrho} must be carried out redundantly on
each node, so that there is minimal speedup; however, since {\bf vofrho} never
requires more than about 10\% of the CPU time, this is not a major handicap.
The greatest challenge in the orbital decomposition is the parallel
implementation of {\bf ortho}.  Good speedup is achieved only for small numbers
of processors; the time required for {\bf ortho} quickly approaches a minimum
due to communication required when doing the parallel matrix multiplications.
A pure spatial decomposition, on the other hand, shows good efficiencies for
those parts of the code which involve computations strictly in real space or
strictly in reciprocal space, but is much less efficient when transforming back
and forth between the two (Figure \ref{fig2}).  The inefficiencies seen in
those parts of the computation carried out purely in reciprocal space, such as
the second subtask of {\bf vofrho}, the second and third subtasks of {\bf
nonlocal}, and the sums over ${\vec g}$ in {\bf nonlocal} and {\bf ortho}, are
due to load balancing problems; the decomposition of the FFT lattice results in
an uneven division of the coefficients for very large numbers of processors, to
the point where one node may have more than twice as many as some others; the
nodes receiving a larger number of coefficients then become a bottleneck.

   Thus both approaches begin to lose efficiency when the number of processors
becomes large enough, but for different reasons.  The spatial decomposition
begins to suffer from load-balancing problems, and more importantly, it loses
speed in the FFTs due to communication overhead.  The orbital decomposition, on
the other hand, reaches a bottleneck due to the communications when
orthonormalizing the electronic orbitals.  The orbital decomposition is
somewhat faster for a given number of nodes, but it limits the number of nodes
which can be applied to the problem to no more than half the number of orbitals
-- and it requires more memory due to redundant storage on different nodes.  By
combining the two approaches, it is possible to balance the decomposition so as
take maximum advantage of the strong points of each approach.  This hybrid
parallel Car-Parrinello (HPCP) algorithm makes it possible to tune the
decomposition for a given problem and a given number of nodes to get the
maximum speedup.  If the number of nodes available is limited, an optimal
decomposition can be determined and used to minimize computational time.  When
the maximum number of nodes under orbital decomposition has been applied, it is
possible to add spatial decomposition to apply an arbitrary number of nodes.
We have found that, in fact, the optimal decomposition is usually {\it not}
purely spatial or purely orbital, but a combination of the two.


\subsection{The hybrid decomposition}
\label{sec:level41}

   The HPCP technique divides the computing nodes into groups; each group is
assigned a subset of the electronic orbitals, and the computations on these
orbitals are further subdivided spatially among the nodes within the group.
The groups are chosen in such a way that the members can communicate with each
other during the computation without interfering with the communications among
members of other groups; they are compartmentalized to eliminate message
contention during most of the computation.  In addition, the nodes are arranged
so that equivalent nodes -- that is, nodes which have been assigned equivalent
subsets of the expansion coefficients -- in different groups can be mapped into
a set of independent rings with as few shared communications links as possible.
In this paper, we concern ourselves only with the details of implementation on
the Intel Paragon and the Touchstone Delta, two multiple instruction, multiple
dataset (MIMD) architectures with a mesh interconnect communication network.
Implementation on other architectures with higher dimensional communication
networks such as the T3D, which uses a 3-D toroid communication interconnect,
or on the iPSC/860 or nCUBE/2, which use the hypercube interconnect, is
straightforward.  For instance, the subgroups chosen on the T3D might be
"planes" of processors within the 3-D lattice.

   With the mesh interconnect, the computer is viewed as a 2-dimensional set of
nodes, each with a connection to 4 neighbors on the North, East, West, and
South, with the exception of those nodes on the edges of the mesh.  For the
purpose of mapping HPCP onto the nodes, the mesh is viewed as a 2-D mesh of 2-D
submeshes.  Figure \ref{fig3} gives a schematic picture of the three types of
decomposition on a mesh computer for an example problem involving 32 orbitals;
a purely orbital decomposition on the left, with each node responsible for all
computations on two orbitals, a purely spatial decomposition on the right,
where each node is responsible for approximately one sixteenth of the work on
all 32 orbitals, and in the center a hybrid decomposition where the $4 \times
4$ mesh is decomposed into a $2 \times 2$ set of $2 \times 2$ submeshes, each
submesh is responsible for 8 orbitals, and each subnode is responsible for
approximately one fourth of the coefficients for those 8 orbitals.  The
operations which were carried out on a single node in the pure orbital
decomposition are now carried out within the submesh by all the subnodes
working in parallel.  Communications which originally took place between
individual nodes now pass between equivalent nodes within the submeshes.  This
has the effect of reducing communication time in the orthonormalization
procedure considerably.

   The next question, then, is how the coefficients will be assigned to the
various subnodes within each submesh.  Since the calculations for each
plane-wave are identical, with the exception of the $\vec g = \vec 0$, it is
not particularly important which node contains which coefficients.  Also, it is
not important which parts of the real-space simulation cell are assigned to
each node.  However, there are several other considerations.  First: the
partition must be chosen so that the parallel FFTs can be done in an efficient
manner.  Second: the number of coefficients assigned to each subnode should be
roughly equal to balance the load. Third, and most important: since we have
chosen to include only the $\Gamma$ point, the coefficients $c^n_{-\vec g}$ are
actually just the complex conjugates of $c^n_{\vec g}$, reducing the amount of
storage required by half; in order to maintain this advantage, the partition
must be chosen so that it is not necessary for different nodes to maintain
consistent values of the coefficient for a given positive/negative plane-wave
pair; that is, the elements of the FFT array in reciprocal space corresponding
to the positive and negative plane-waves {\it must reside in the same node's
memory}.

   While it is certainly possible to implement true 3-D parallel FFTs, in the
case of a multi-dimensional FFT it is simpler and usually more efficient
\cite{Gupta93,Marinescu92,Angelopoulos93} to implement the 3-D FFT as a series
of 1-D FFTs in the $x$, $y$, and $z$ directions combined with data
transpositions (Figure \ref{fig4}).  We have chosen to partition the data so
that in reciprocal space, the entries for the $z$ dimension are stored
contiguously in local memory, and the $x$ and $y$ dimensions are decomposed
across the nodes; thus, each node has a set of one-dimensional columns on which
to work when doing the one-dimensional FFTs.  The exact nature of this
decomposition is determined by the second and third requirements mentioned
above.  A straightforward partitioning on a $4x4$ submesh might be done as in
Figure \ref{fig5}(a); this would make data transpositions quite simple, but
would lead to major load-imbalance problems due to the fact that most of the
entries in the FFT array in reciprocal space are actually zero.  Only those
plane-waves within the cutoff energy are actually used in the calculation; they
fill only a relatively small region (Figure \ref{fig6}) within the actual FFT
array.  It is this {\it sphere of active plane waves} within the FFT array
which must be evenly divided, if the computations in reciprocal space are to be
evenly divided among nodes.  So the simple partition is discarded in favor of
an {\it interleaved} partition (Figure \ref{fig5}(b)), which results in a
roughly equal division of the active plane-waves among the subnodes.

   When doing a standard one-dimensional FFT of an array $ f $ with length $ L
$ indexed from $ f_0 $ to $ f_{L-1} $, the values of $ f_n $ in the array in
real space correspond to values of some function $ f $ for equally-spaced
values of some variable $ x $, arranged in ascending order.  When the array is
transformed into reciprocal space, we are left with a new array $ F $ whose
entries $ F_n $ correspond to the intensities of various frequencies $ n \omega
$ in a Fourier expansion of the function $ f $.  They are not, however,
arranged in simple ascending order; rather, $ F_0 = F(0) $, $ F_n = F( n \omega
) $ for $ 0 < n < {L \over 2} $, and $F_n = F( (n - L) \omega )$ for $ {L \over
2} \leq n < L $.  Hence negative frequency $ - m \omega $ maps to array
location $ F_{L-m} $.  We must account for the fact that this will be the case
in all three dimensions of our FFT array; positive plane wave $ ( n_x, n_y, n_z
) $ will map to negative plane wave $ ( m_x, m_y, m_z ) $ in a rather
complicated manner depending upon the signs and values of $n_x$, $n_y$, and
$n_z$; for instance if we use the simple interleaved partition suggested in
Figure \ref{fig5}(b), plane wave $( 1, 2, 3 )$ would be stored in the local
memory of node 9, but its negative $( -1, -2, -3 )$ would map to $( 15, 14, 13
)$ and end up in the local memory of node eleven.  Either it is necessary to
double the storage for coefficients and determine a way to maintain
coefficients for positive and negative plane waves as complex conjugates, or it
is necessary to choose the data decomposition so that they are both stored in
the same node's memory.  The problem is simplified considerably if one looks at
it in terms of the values of the plane waves rather than the partitioning of
the FFT array itself.  If node $p$ is at location $(i,j)$ in an $R \times C$
submesh, then those plane waves with $ | n_x | {\rm mod} C = j $ {\it and} $ |
n_y | {\rm mod} R = i $ should be assigned to $p$.  Assigning coefficients to
nodes according to the absolute value of $n_x$ and $n_y$ forces the
decomposition to satisfy the third condition, since {\it all} values in the $z$
dimension are known to be in local memory already.  A simple example of the
final decomposition is shown in Figure \ref{fig5}(c).


\subsection{The FFT calculations}
\label{sec:level42}

   Once the partition has been made, the FFT is straightforward.  For
simplicity, we consider only the transform from reciprocal space to real space;
the reverse is analogous.  Each node in the submesh begins with a subset of
columns in the $z$ dimension; it performs the one dimensional FFTs on these,
then begins preparing to transpose the data so that it will have a subset of
columns in the $y$ dimension.  The data is packed into a set of contiguous
buffers so that the data which will remain on the node is placed in the "first"
buffer, the data which needs to be transmitted to the node immediately
"beneath" it goes in the next buffer, the data for the node "beneath" that one
goes in still the next buffer, and so on; when the "bottom" of the mesh is
reached, it wraps back to the top row in a toroid-fashion.  Once the data is
packed, a series of messages are passed along each column of subnodes; after
the first message, each node has the data it requires from the node directly
above it, as well as the data which was already in local memory.  Each node
retains the buffer which was meant for it, passes on the buffers which were
meant for the nodes "below" it, and receives a new set of buffers from the node
"above" it.  After all the buffers have been sent, each node will have the data
necessary to reconstruct the FFT array, with a subset of the columns in the $y$
dimension in local memory.

   This procedure is shown schematically in Figure \ref{fig7}.  The buffers $
B_{ij} $ are contiguous in memory so that the messages, once packed, may be
sent without any further data movement.  Buffer $ B_{ij} $ is the data on node
$i$ which must be transmitted to node $j$.  At each iteration, only the shaded
buffers are transmitted to the next node.    We use this store-and-forward
technique because the software "bandwidth" on the Touchstone Delta, i.e. the
amount of data which can be transferred from local memory out to the message
network per unit time, is sufficiently close to the hardware bandwidth that it
is possible to swamp the message backplane and degrade communications.  The
store-and-forward technique reduces all communications to near-neighbor
messages, eliminating this possibility.  On the Paragon, however, the hardware
bandwidth is almost an order of magnitude greater than the speed with which any
particular node can move data from its local memory out onto the network, so it
is almost impossible to swamp the backplane; it may be that direct messages
will be faster under these conditions.

   Once the FFT array is reassembled, each node performs the appropriate one
dimensional FFTs on the subset of the columns in the $y$ dimension in its local
memory.  When these are done, the nodes again pack a set of message buffers,
but this time the messages will be passed {\it horizontally} along each row of
the submesh in order to transpose the $y$ and $x$ coordinates; when the
messages have been passed and the buffers unpacked, each node will have a
subset of the columns in the $x$ dimension in local memory.  Appropriate FFTs
are performed, and the three dimensional parallel FFT is complete.  The FFT
back to reciprocal space simply performs the same operations in reverse order,
doing the $x$ FFTs first, transposing $x$ and y, doing the $y$ FFTs,
transposing $y$ and z, and finally doing the $z$ FFTs.  An overall schematic of
the FFT showing the data flow during the transpose operations is shown in
Figure \ref{fig8}.


\subsection{Global summations}
\label{sec:level43}

   Global summation is a parallel operation whereby the sum of the values
stored in some variable or variables on different nodes is calculated.

   In the HPCP code, three different global summations are needed.  Referring
to the hybrid decomposition diagram in Figure \ref{fig3}, those are: (1) a
standard global summation, where all the nodes, zero through fifteen,
contribute to the sum; (2) a global summation {\it over all nodes within a
submesh}, for instance, only summing up values of the variable stored on nodes
two, three, six, and seven; and (3) a global summation {\it over equivalent
nodes in all the submeshes}, for instance over nodes five, seven, thirteen, and
fifteen.

   Standard library calls on most parallel computers can do the first type of
summation.  The other two types of sums need to be specially coded.  We have
used as a basis for our global summations the technique suggested by
Littlefield and van de Geijn \cite{Littlefield93}, modified for the specific
circumstances.  The general technique involves independent sums along the rows
(or columns) of the mesh, leaving the first node in the row (or column) with
the sum of the values along the row (or column).  Then a summation is done
along the first column (or row), leaving the node in the $(0,0)$ position with
the result.  The communications pattern is then reversed, as the result is
broadcast out along the first column (or row) and then independently along each
row (or column).  If properly done, there is no message contention and the time
required for completion scales as $log_2P$, with $P$ the number of nodes.
Doing this {\it within a submesh} is straightforward; there is no possibility
of message contention, since no node within the submesh ever needs to send a
message outside the boundaries of the submesh.

   Doing summations over equivalent nodes, however, presents a problem; there
will inevitably be message contention for large enough numbers of processors.
It can be minimized, however, by taking advantage of the fact that it is
arbitrary whether the first phase of the summation is done along the rows or
the columns.  If all sets of equivalent nodes do their summations with
identical ordering of messages, a great deal of contention will be introduced
(Figure \ref{fig9}(a)).  If, on the other hand, every other set, selected
according to a "checkerboard" coloring, does the first stage along the {\it
rows}, message contention is reduced by half (Figure \ref{fig9}(b)).


\subsection{Matrix multiplication}
\label{sec:level44}

   In order to perform the iterative orthonormalization efficiently, it is
necessary to have an efficient parallel matrix multiplication routine.  The
matrix multiplication used for the pure orbital decomposition can be extended
to make efficient use of the hybrid node layout, increasing the effective
bandwidth by roughly the square root of the number of nodes in each submesh.
To illustrate this, it is necessary to review the way in which the matrix
multiplication was implemented in the earlier, pure orbital decomposition code
\cite{Wiggs94}.

   First, a "ring" was mapped into the underlying communication topology so
that each node had a "neighbor" on its "left" and its "right," to which it had
a direct, unshared communications link.  The coefficient array was divided up
among the members of this ring so that each node had an equal number of rows.
Doing the large matrix multiplications when evaluating Eqs. (\ref{eq25}) and
(\ref{eq26}) required that each node use the rows of the coefficient matrices
stored in its local memory to calculate a subblock of the result matrix, pass
its rows along to the next node in the ring, and receive the next set of rows
from the previous node in the ring.  The algorithm was described in more
detail, with schematics, in \cite{Wiggs94}.

   This parallel matrix multiply was free of contention and maintained load
balancing, but its efficiency was greatly limited by the available bandwidth
between nodes.  Its performance dropped off rapidly as the number of nodes
increased, due to time spent waiting for the messages to get around the ring.
As an example, results of calculations on a 64 Si atom system on the Paragon
and the Delta are shown in Figure \ref{fig10}.  The time for each is well
described by the relation $t = a + {b \over P}$, where $P$ is the number of
nodes and $a$ is the time spent in communications.  This baseline is
independent of the number of nodes due to the fact that as the number of nodes
increases, the number of messages also increases but the {\it size} of the
messages decreases in exact proportion; the total amount of data that each node
is required to send and receive is independent of the number of nodes.  The
baseline for the Paragon is less than half that of the Delta, owing to the
former's greater node-to-node communication bandwidth.

   A hybrid parallel matrix multiply greatly alleviates this communication
bottleneck.  Each node within a submesh has a subset of the coefficients for
some subset of the orbitals.  If equivalent processors from the submeshes are
mapped into independent rings, each can do a matrix multiply with its own
submatrix; the final result matrix can be obtained by doing global sums within
each submesh of the result matrices produced by each node (Figure \ref{fig11}).


\subsection{Parallel algorithm for {\bf ortho}}
\label{sec:level45}

   Once the parallel matrix multiplication has been implemented, the ortho
routine is relatively simple.  The arrays {\bf b} and ${\bf X^{(0)}}$  (Eqs.
\ref{eq25} and \ref{eq26}) are calculated by multiplying the coefficient
matrices by their transposes; the data is already properly laid out for the
matrix multiplication technique described above.  The array ${\bf X^{(0)}}$ is
constructed according to Eq. \ref{eq25} by multiplying the new matrix of
coefficients for time $t + \delta t$, which are not orthonormalized, by its own
transpose.  The array {\bf b} (Eq. \ref{eq26}) is constructed by multiplying
this matrix by the transpose of the matrix of coefficients for the current time
$t$, which are already orthonormalized.  The iteration involves similar matrix
multiplications of arrays ${\bf X^{(0)}}$ and {\bf b}, or linear combinations
thereof, by the transposes of other linear combinations; thus the same matrix
multiplication technique can be applied to the calculations within the
iteration loop, without need for any matrix transpositions.  These matrix
multiplications within the iteration loop are carried out redundantly on each
set of equivalent nodes, since each node in the submesh will need a copy of the
result matrix ${\bf X^{(k)}}$ to correct the new array of coefficients;
however, since the amount of work required for these is  $O(N^3)$ rather than
the $O(N^2 M)$ required to calculate {\bf b} and ${\bf X^{(0)}}$ ($M >> N$),
the redundant calculation has little effect on the overall timing of {\bf
ortho}.

   The {\bf ortho} routine has a decided advantage over the GS technique when
doing large problems and decomposing by orbitals; the amount of data
communication it requires does not grow with the number of nodes, whereas the
GS technique's communication requirements increase as the number of orbitals
per node decreases.  In the GS procedure the nodes do a sequential loop over
all orbitals; the node which "owns" the orbital normalizes it, then {\it
broadcasts} it to all the other nodes, who then orthogonalize to it any of
those orbitals they "own" which have not already been so treated.  As such, the
algorithm requires $N$ broadcasts, each of which requires $log_2P$ messages to
be sent if there are $P$ nodes.  For a problem with $N$ orbitals, $M$
coefficients, and $P$ nodes, the communication requirements for GS are then:
$O( NMlog_2P )$.  The {\bf ortho} routine, on the other hand, requires 3 large
matrix communications like those described in Section \ref{sec:level44}, each
of which requires $P$ messages of length $M {N \over P}$ (the time for the
smaller $N \times N$ matrix multiplications is swamped by these), implying O(
NM ) communications regardless of the value of $P$.  Of course, in the hybrid
implementation, $P$ refers to the number of {\it submeshes}, as this determines
how many messages must be sent.  A purely spatial decomposition would not
require any of the communication described here -- only a series of global sums
to determine the orbitals' overlaps.


\subsection{Parallel algorithm for {\bf rhoofr}}
\label{sec:level46}

   The algorithm for {\bf rhoofr}, which calculates the total electronic
density within the simulation cell, is quite simple.  Each submesh loops over
the orbitals assigned to it, placing the coefficients for each into the proper
locations in the FFT array and doing inverse FFTs to get the value of the
electronic wavefunction in real space; the entries are then squared, scaled by
the appropriate occupation number $f_n$, and accumulated in the array {\bf
rhoe}.  When this loop is completed, independent global sums of {\bf rhoe} are
calculated {\it over equivalent nodes}, leaving each node with the total
electronic density in that subregion of the simulation cell for which it is
responsible.



%
-

\section{Some Performance Numbers}
\label{sec:level5}

   The HPCP code makes it possible to "tune" the decomposition for a particular
problem and given number of nodes to minimize the time required for the
calculation.  To demonstrate this, we describe below several test runs done on
both the Delta and Paragon computers.  The calculations were done on systems of
Si atoms in the crystalline arrangement, using s and p non-locality BHS
pseudopotentials with the KB factorization, and an energy cutoff for the plane
wave expansion of 12 Ry.  Figure \ref{fig12} shows timings from calculations
done using 64 nodes on the Touchstone Delta, starting with a purely orbital
decomposition (1 node per submesh) and increasing the number of nodes per
submesh up to 64 nodes, giving a purely orbital decomposition.  The major
subroutines described above were timed separately.  The behavior of {\bf
rhoofr}, {\bf local}, and {\bf ortho} in particular are of interest and are
described in more detail below.

   {\bf Local} represents the time required to apply the local part of the
pseudopotential; it requires two FFTs for each orbital.  For each orbital in
turn, the coefficients are placed into the FFT array, an inverse FFT is
performed to get the value of the wavefunction in real space, the real space
wavefunction in multiplied by the potential calculated in {\bf vofrho}, and a
forward FFT is done, yielding the values of the "local" force on the electronic
coefficients.  {\bf Rhoofr} requires only one FFT per orbital, as described in
Section \ref{sec:level46}.  The rapid increase in time required for these two
routines is due to the loss of efficiency in the parallel FFTs.  The actual
amount of computation per node does not change, but the communication time
increases as more nodes are involved in each FFT calculation.  It is impossible
for any parallel FFT to be more efficient than simply dividing up the FFTs and
doing them independently on separate nodes as is the case in the orbital
decomposition.  Clearly, for these two tasks, the orbital decomposition is
superior.

   On the other hand, the timing in {\bf ortho} improves rapidly as the number
of nodes per submesh increases.  The reason is that the amount of communication
required for the matrix multiplications decreases as the number of submeshes
decreases.  For a purely spatial decomposition, no communication is required to
do the multiplication of the submatrices; the only communication required is
the final global sum to get the total values in the result matrix.  For a
hybrid decomposition, as the number of submesh nodes increases, the number of
submeshes decreases, and the effective bandwidth increases roughly as the
square root of the number of nodes in each submesh.  This behavior is
illustrated in Figure \ref{fig13} for a system with $N$ orbitals and $M$
coefficients.  This analysis is born out by the actual timings gathered; when
the number of nodes is fixed at 64 and the number of subnodes per submesh
increases from one to four, the time required for ortho drops from 5.05 seconds
to 2.18 seconds -- slightly better than the factor of two which was expected.
As the number of nodes per submesh is increased further, the time spent in
ortho begins to be dominated by the communications required for the final
global sum in each matrix multiplication.  No improvement in timing is obtained
beyond sixteen subnodes.

   Another advantage of the HPCP algorithm is that it allows more nodes to be
applied to the problem than could be done with the purely orbital
decomposition.  For example, a system of 64 Si atoms has 128 orbitals; the
largest number of nodes which could be applied to this in the pure orbital
decomposition is 64.  Using the HPCP code, the number can then be expanded;
instead of 64 nodes we can, for example, use 64 submeshes, each with as many
subnodes as we wish.  In this case, the effective bandwidth for the matrix
multiplications will again increase roughly as the square of the number of
processors, though for a different reason.  If we increase the computational
mesh from an $8 \times 8$ mesh to a $16 \times 16$ mesh divided into $2 \times
2$ submeshes, the communication pattern among equivalent subnodes is identical;
each subnode will see itself as a member of an $8 \times 8$ mesh made up of
itself and the equivalent subnodes in the other submeshes.  The number of
messages required when doing the matrix multiplications remains the same, and
each subnode will be sharing a communication link with one other (cf. Figure
\ref{fig13}), but the messages will be roughly one-fourth the {\it size} of
those in the $8 \times 8$ mesh, so the overall speed will increase by roughly a
factor of two.  Again, the actual timings gathered bear this out; the time for
{\bf ortho} on an $8 \times 8$ mesh is 5.05 seconds, and on a $16 \times 16$
mesh divided up into $2 \times 2$ submeshes, it is 2.05 seconds.

   From Figure \ref{fig12}, it is apparent that most of the speedup obtained in
{\bf ortho} by increasing the number of subnodes is achieved before the time
spent in {\bf local} and {\bf rhoofr} increases significantly, with a minimum
for the overall time when the number of nodes per submesh is four or eight.
The optimal decomposition is one which is neither purely spatial nor purely
orbital, but a combination of the two.  Figure \ref{fig14} shows the total time
per iteration of the CP algorithm in a simulation of 64 Si atoms.  It is
possible to outperform a highly optimized sequential version of the CP
algorithm running on the Cray Y-MP using only 64 nodes of the Delta, and to
almost match the times on the Cray C90 using 256 nodes, even for a problem as
small as 64 Si atoms.  Figure \ref{fig15} shows a similar comparison for a
larger system of 128 Si atoms.  As the size of the problem increases, the
parallel algorithm does even better; it is possible to outperform the C90 using
256 nodes on the Delta, decomposed as a $4 \times 4$ mesh of $4 \times 4$
submeshes.

   The main motivation for parallel implementations of the CP algorithm is to
open the possibility of simulating systems with more atoms than are possible
with existing and foreseeable sequential computers.  As such, most groups that
have implemented CP or similar electronic-structure calculations on a parallel
supercomputer have reported timings for a few large systems.  Both Nelson
et.al. \cite{Nelson93} and Brommer et.al. \cite{Brommer93} report timings for a
system of 512 Si atoms; in order to make a comparison, we have done likewise on
the Touchstone Delta.  Nelson et.al. implemented their code on an nCUBE 2 with
1024 nodes located at the San Diego Supercomputer Center; the calculation they
timed was 512 Si atoms with one ${\vec k}$-point and an 8-Ry energy cutoff;
their time per iteration was $\sim$12,000 seconds \cite{Nelson93}.  Brommer
et.al. implemented their code on the Connection Machine CM-2, a Single
Instruction, Multiple Data parallel computer.  Their CM-2 was equipped with
65,536 bit-serial processors and 2048 64-bit floating point units.  Their
simulation was done on 512 bulk Si atoms with a 12-Ry energy cutoff.
Estimated by reading from a log-log scale graph (Fig. 9 in \cite{Brommer93}),
the time per iteration was $\sim$3,000 seconds.  For our calculation, we used
512 nodes of the Touchstone Delta.  The nodes were mapped as a $4 \times 8$
mesh of $4 \times 4$ submeshes, and the time per iteration was 250 seconds.  A
concise list of times is given in Table \ref{table2}.  It must be stressed that
the other groups may not have implemented precisely the same algorithm; for
instance Nelson et. al. actually implemented the first-order equation of motion
approach of Williams and Soler \cite{Williams87} which, while very similar to
CP, is not identical.


\section{Conclusions}
\label{sec:level6}

   The implementation of the Car-Parrinello algorithm on scaleable massively
parallel computers permits simulations of systems too large to simulate on
conventional vector supercomputers.  Hybrid spatial and orbital decomposition
of the problem makes it possible to achieve substantially higher throughput
with a given number of processing nodes; neither a purely spatial nor a purely
orbital decomposition gives optimal speedup.


\section{Acknowledgments}
\label{sec:levelack}

   We thank Prof. Roberto Car for making his CRAY code available to us.  We are
indebted to Dr. Arthur Smith for various helpful discussions.  Thanks also to
Dr. Rick Kendall and Dr. Rik Littlefield of Battelle Pacific Northwest Labs for
various helpful discussions on parallel implementations in particular on the
Delta computer, and to Dr. Larry Snyder and his research group at the Computer
Science and Engineering Department of the University of Washington for various
helpful discussions and suggestions on parallel computing in general.  And
finally, thanks to Greg Mills for his suggestions which led to the data
decomposition used for the parallel FFTs.

   This research has been supported by the Department of Energy's Computational
Science Graduate Fellowship Program (JW) and by NSF-CARM grant CHE-9217294.
Computer time was made available on the Delta by PNL-Battelle in Richland, WA,
on the Paragon at Oak Ridge National Labs through the National Energy Research
Supercomputer Center at Lawrence Livermore National Labs, and on the Paragon
and CRAY Y-MP8/684 by the San Diego Supercomputer Center.




\vfill\eject

\begin{figure}
\caption{CPU time for major tasks in a pure orbital decomposition parallel
Car-Parrinello simulation of 32 Si atoms at a 12 Ry cutoff energy.  Efficiently
parallel tasks should require approximately half the processing time when the
number of nodes is doubled.  Examples of this are {\bf rhoofr} and {\bf
local}.}
\label{fig1}
\end{figure}


\begin{figure}
\caption{CPU time for major tasks in a purely spatial decomposition parallel
Car-Parrinello simulation of 32 Si atoms at a 12 Ry cutoff energy.  Efficiently
parallel tasks should require approximately half the processing time when the
number of nodes is doubled.  Examples of this are {\bf nonlocal} and {\bf
ortho}.}
\label{fig2}
\end{figure}


\begin{figure}
\caption{Three possible data decompositions for a 16 Si atom simulation on 16
nodes.  In the purely orbital, each node is responsible for 2 out of 32
orbitals.  In the purely spatial, each node is responsible for part of the
computation on all 32 orbitals.  In the hybrid, each node is a member of a
submesh, responsible for part of the computation on 8 of the 32 orbitals.}
\label{fig3}
\end{figure}


\begin{figure}
\caption{Data-transpose algorithm for a 2-dimensional parallel FFT.  First
one-dimensional FFTs are applied in the x-direction, then the matrix is
transposed, then one-dimensional FFTs are applied in the y-direction.  All FFTs
are computed strictly in local memory.  Communication is done only in the
matrix transpose.}
\label{fig4}
\end{figure}


\begin{figure}
\caption{(a)  Simple, inefficient partitioning of the 3D FFT array across
nodes; each node gets 16 columns in the z-direction, but some nodes will have
few or no "active" plane waves assigned to them.  (b)  The interleaved
partitioning assures that each node will get a reasonable share of the "active"
plane waves, but does not address the problem of memory locations for
positive-negative plane wave pairs.  (c)  The final partitioning.}
\label{fig5}
\end{figure}


\begin{figure}
\caption{An $x-y$ projection of the FFT array showing the "active" elements in
the array.  Only those ${\vec g}$ falling within a cutoff radius determined by
$E_{cut}$ have coefficients with non-zero values; they take up the 8 "corners"
of the array cell.}
\label{fig6}
\end{figure}


\begin{figure}
\caption{Each node packs message buffers in memory so that it may successively
send and receive smaller messages, each time receiving part of the data
necessary for it to continue the next step in the algorithm, and also storing
and forwarding to other nodes data they require, withouth introducing a great
deal of message contention.}
\label{fig7}
\end{figure}


\begin{figure}
\caption{The parallel 3D FFT algorithm; 1D FFTs are applied successively in
each dimension, with transpositions of the data array between.  The individual
rows and columns of nodes may carry out the transpositions of their local data
independently, greatly increasing bandwidth and efficiency.}
\label{fig8}
\end{figure}


\begin{figure}
\caption{(a) All sets of nodes do global sums in the same order; first along
columns, then along rows.  One set of equivalent nodes is highlighted for
emphasis.  In each place where two parallel message indicators overlap, there
is contention for a communication link.  (b) If the nodes within a submesh are
colored "red-black" like the pattern of a checkerboard, and the "red" nodes do
their sums first along the rows while the "black" nodes do theirs first along
the columns, the message contention is reduced by half.}
\label{fig9}
\end{figure}


\begin{figure}
\caption{CPU time spent in the orthogonalization routine on the Delta and
Paragon computers in a simulation of 64 Si atoms using pure orbital
decomposition.  As the number of nodes used in the computations is increased
the computational time decreases but the communications time stays the same.
The lower baseline for the Paragon curve demonstrates the higher communication
bandwidth, which reduces the time spent in communications.}
\label{fig10}
\end{figure}


\begin{figure}
\caption{The hybrid parallel matrix multiplication algorithm.  Each node
carries out an independent matrix multiplication of the parts of the matrices
in its local memory, then passes its subblock of the second matrix to the next
node in the ring of equivalent nodes in the other submeshes; arrows indicate
direction of data movement between nodes of each ring.  The shaded blocks in
the result matrix indicate the part being calculated at each iteration in
submesh 1; the shaded blocks in the matrices being multiplied indicate the part
of these matrices resident in memory on each subnode of submesh 1 at each
iteration.  {\it Each} node in the submesh keeps a copy of the indicated strip
of the result matrix; when the independent matrix multiplications are done, a
global summation of the values in that strip of the result matrix is carried
out within each submesh, leaving each subnode with the final result.}
\label{fig11}
\end{figure}


\begin{figure}
\caption{Time spent on the various subroutines of the HPCP algorithm vs. the
number of nodes in each submesh, in a simulation of 64 Si atoms.  The timings
are from the Delta, using 64 nodes total.  One node per submesh corresponds to
a pure orbital decomposition, 64 nodes per submesh corresponds to a pure
spatial decomposition.}
\label{fig12}
\end{figure}


\begin{figure}
\caption{Breakdown of communications requirements and effective bandwidth for
{\bf ortho} in a purely orbital and a hybrid decomposition.  The hybrid
decomposition increases the speed of the algorithm by reducing the number of
messages required by $1/4$ while only decreasing the bandwidth available to
each node by $1/2$.}
\label{fig13}
\end{figure}


\begin{figure}
\caption{Total time per iteration for a system of 64 Si atoms on four different
architectures.  The two horizontal lines give timings for the Cray Y-MP and the
Cray C90; the curves represent timings for different numbers of nodes on the
Touchstone Delta and the Intel Paragon. Timings were done with 64, 128, and 256
nodes of the Toucshtone Delta and with 64 nodes of the Intel Paragon.  The
curves indicate how the time per iteration changes with the decomposition; the
$x$ axis is the $log_2$ of the number of nodes per submesh.}
\label{fig14}
\end{figure}


\begin{figure}
\caption{Total time per iteration for a system of 128 Si atoms on two different
architectures.  The horizontal line gives the time per iteration on the C90;
the curves represent timings for 128 and 256 nodes of the Toucshtone Delta.
Again, the $x$ axis is the $log_2$ of the number of nodes per submesh.}
\label{fig15}
\end{figure}

\vfill\eject


\widetext
\begin{table}
\caption{Operations in the Car-Parrinello algorithm for N orbitals (indexed
with $n$) and M plane waves (indexed with $\vec g$). \label{table1}}

\vskip 0.5 true cm

\begin{tabular}{lcccccc}
Task & Operations \\ \tableline

rhoofr:& $c^n_{\vec g} \ \ {\buildrel{\rm FFT} \over \longrightarrow} \ \
\Psi_n(\vec r)$ \\

& $\rho(\vec r) \ = \ \sum^N_n \ 2 \ |\Psi_n(\vec r)|^2 $ \\

\\ vofrho:& $\rho(\vec r) \ \ {\buildrel{\rm FFT} \over \longrightarrow} \ \
\rho(\vec g)$ \\

& $\tilde v (\vec g) \ = \ {{4 \pi} \over {|\vec g|^2}} \ \rho(\vec g) \ + \
v_{loc}(\vec g)$ \\

& $\tilde v(\vec g)\ \ {\buildrel{\rm FFT}\over \longrightarrow}\ \ \tilde
v(\vec r)$ \\

& $ v(\vec r) \ = \ \tilde v (\vec r) \ + \ \mu_{xc}(\rho(\vec r))$ \\

\\ nonlocal:& $ F_l^{In} = \sum^M_{\vec g} { u_l(\vec g) e^{i \vec g \cdot
{\vec R}_I}  c^n_{\vec g} }$ \\

& ${{\partial {\tilde v}^{l}_{nl}(\vec g) } \over {\partial c^n_{\vec g}}} = 2
{ {\sum_I e^{-i \vec g \cdot {\vec R}_I} u_l(\vec g) F_l^{In}} \over { W_l } }
$ \\

& ${{\partial  E } \over {\partial c^n_{\vec g}}} \ = \ {{\partial \tilde E }
\over {\partial c^n_{\vec g}}}\ +  \sum^{l_{m}}_{l=0} {{\partial {\tilde
v}^{l}_{nl}(\vec g) } \over {\partial c^n_{\vec g}}}$ \\

\\ local:& $c^n_{\vec g} \ \ {\buildrel{\rm FFT} \over \longrightarrow} \ \
\Psi_n(\vec r)$ \\

&$2 \Psi_n(\vec r) \ v(\vec r)\  \ {\buildrel{\rm FFT} \over \longrightarrow} \
\ {{\partial \tilde E } \over {\partial c^n_{\vec g}}} $ \\

\\ loop:& $ \bar c^n_{\vec g}(t+\delta t) = - c^n_{\vec g}(t-\delta t) + 2
c^n_{\vec g}(t)  -  {{(\delta t)^2} \over \mu} \ {{\partial E } \over {\partial
c^n_{\vec g}}}$ \\

\\ ortho:& $2X^{(0)}_{nm} = \delta_{nm}  - \sum^M_{\vec g} \ \bar c^{n*}_{\vec
g}(t + \Delta t) \  \bar c^m_{\vec g}(t + \Delta t)$ \\

& $b_{nm}  = \delta_{nm}  -  \sum^M_{\vec g} \ c^{n*}_{\vec g}(t) \ \bar
c^m_{\vec g}(t+\Delta t)$ \\

& $X^{(k)} = X^{(0)}  +  X^{(k-1)}  b +  b^T  X^{(k-1)}  - X^{(k-1)^2}$ \\

&$c^n_{\vec g}(t+\Delta t) = \bar c^n_{\vec g}(t+\Delta t) + \sum^N_{m=1}
X^{(k)}_{nm} c^m_{\vec g}(t)$

\end{tabular}
\end{table}

\vfill\eject

\widetext
\begin{table}
\caption{Timing comparisons of various simulations of 512 Si atoms.
\label{table2}}

\vskip 0.25 true cm

\begin{tabular}{cccccc}
 Reference: & Computer & Nodes & Decomposition & $E_{cut}$ (Ry) & sec/step \\
\tableline
 Nelson et. al. \cite{Nelson93} & nCUBE/2 & 1024 & Spatial & 8.0 & 11,677 \\
 Brommer et. al. \cite{Brommer93} & CM-2 & 64K & Spatial & 12.0 & 3,000 \\
 Present work & Delta & 512 & Hybrid & 12.0 & 250 \\
\end{tabular}
\end{table}

\vfill\eject

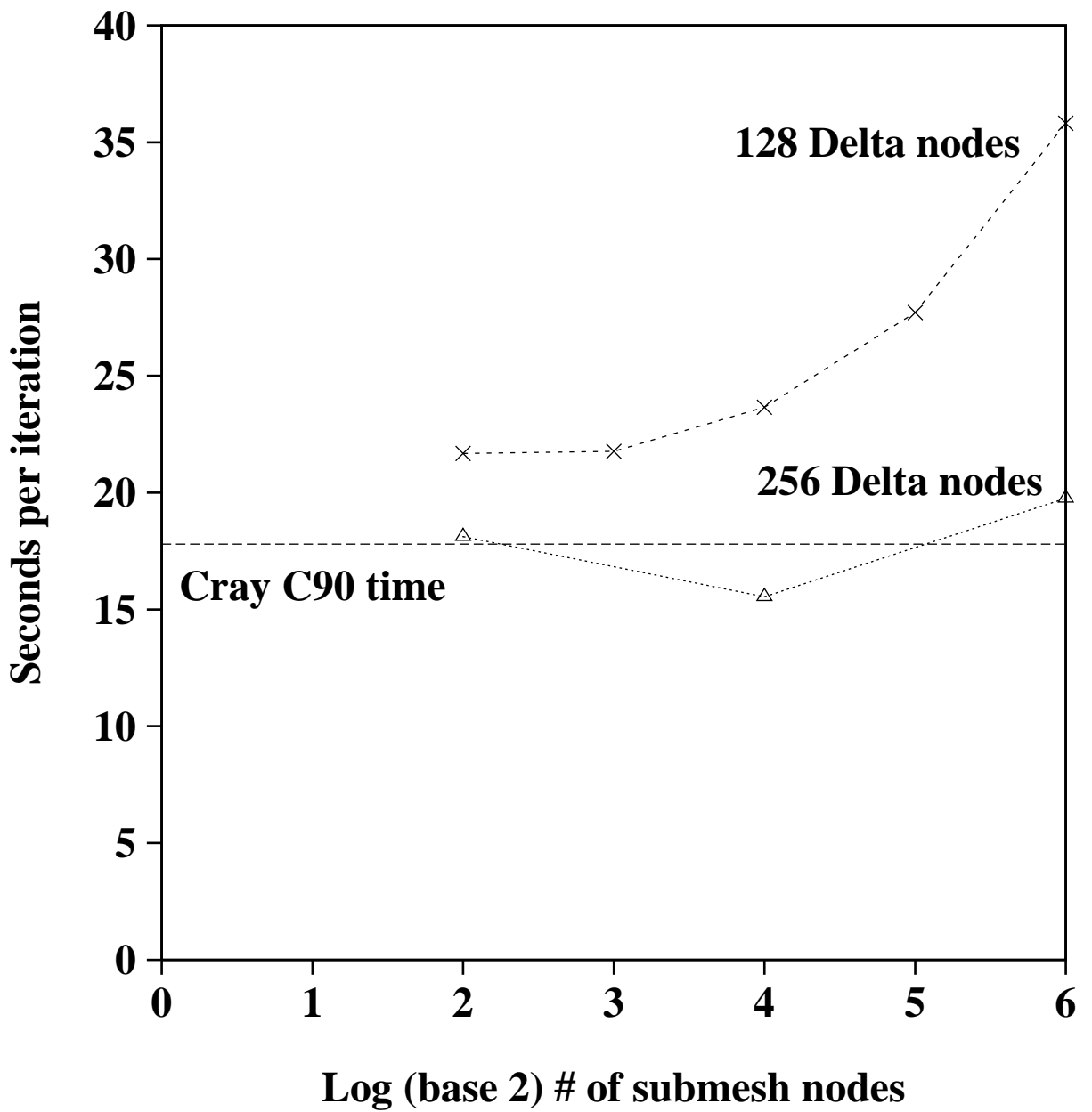
\end{document}